\documentclass[12pt]{article}

\newcommand{\x}{arXiv:}
\newcommand{\m}{\mathrm}
\newcommand{\be}{\begin{equation}}
\newcommand{\ee}{\end{equation}}
\newcommand{\ba}{\begin{eqnarray}}
\newcommand{\ea}{\end{eqnarray}}

\usepackage{graphicx}
\usepackage{amssymb}
\usepackage{amsmath}
\usepackage[T1]{fontenc} 
\usepackage[ansinew]{inputenc} 
\usepackage[nosort]{cite}
\newcommand{\inbar}{\vrule height1.57ex width.4pt depth0pt}
\newcommand{\SW}{\relax{\hbox{$\ \inbar\kern-.285em{\rm S}$}}}

\begin{document}
\thispagestyle{empty}
\begin{center}

\null \vskip-1truecm \vskip2truecm

{\Large{\bf \textsf{Viscosity vs. Vorticity in the Quark-Gluon Plasma}}}

{\Large{\bf \textsf{}}}

{\large{\bf \textsf{}}}

\vskip1truecm

{\large \textsf{Brett McInnes
}}

\vskip0.1truecm

\textsf{\\ National
  University of Singapore}
  \vskip1.2truecm
\textsf{email: matmcinn@nus.edu.sg}\\

\end{center}
\vskip1truecm \centerline{\textsf{ABSTRACT}} \baselineskip=15pt
\medskip

We show that, in a holographic or gauge-gravity approach to the study of the Quark-Gluon Plasma, the specific entropy density of the plasma produced in a peripheral heavy-ion collision can be strongly suppressed relative to its value for central collisions at the same impact energy; this is connected with the strong vorticity fields which are now known to characterize the peripheral case. This implies, within the holographic model, that the shear viscosity of ``peripheral plasmas'' may be considerably lower than that of the corresponding ``central plasmas'', with potentially important consequences for other aspects of QGP physics, such as the centrality dependence of jet quenching.

\newpage
\addtocounter{section}{1}
\section* {\large{\textsf{1. Centrality Dependence of Viscosity }}}
The Quark-Gluon Plasmas (QGP) thought to be produced in collisions of heavy ions \cite{kn:plosk,kn:oet} are currently objects of intense interest. A major theme here is the \emph{centrality dependence} of various properties of the QGP: that is, one compares the outcomes of peripheral collisions, at a specific range of impact parameters, with those of their central counterparts.

One of the most important properties of the QGP is its \emph{viscosity}. As the various beam energy scan experiments (see for example \cite{kn:tlusty}) probe the quark matter phase diagram, one hopes to answer various fundamental questions: for example, does the QGP become more liquid-like or more gas-like at high values of the baryonic chemical potential \cite{kn:yiyin}? The behaviour of the viscosity as collision parameters vary will surely play an important role in answering such questions.

Combining these two concerns, we ask: what is the nature of the \emph{centrality dependence of the QGP viscosity}? Where should we expect this dependence to appear in the data?

To take a specific example of great current interest: \emph{jet quenching} \cite{kn:jet,kn:jetagain,kn:nattrass} is the tendency of the strongly coupled plasma to reduce the energy, and to broaden the transverse momentum $p_T$, of hard partons propagating through it. The centrality dependence of this phenomenon is beginning to be explored (see for example the STAR collaboration results \cite{kn:beamenergyjet} and also the investigations of the ALICE collaboration at the LHC \cite{kn:alicemod}); comparisons are made between the various numbers of high-$p_T$ hadrons produced in central collisions with those produced in peripheral collisions.

Any theoretical analysis of these observations must assume, among many other things, that one has a good understanding of the centrality dependence of the QGP viscosity. For one certainly expects the viscosity of the plasma to affect jet quenching: in particular, both are connected with the entropy density of the plasma.

Peripheral plasmas are created in an environment and with properties that are quite unique. In particular, they can exhibit large \emph{vorticities}, a fact long predicted \cite{kn:liang,kn:bec,kn:huang}, and recently confirmed experimentally \cite{kn:STARcoll,kn:STARcoll2,kn:STARcoll3,kn:STARcoll4} by the STAR collaboration \cite{kn:STAR} at RHIC; and new ways of probing this phenomenon are being proposed (see for example \cite{kn:bhatt}). Our question now takes a more concrete form: can these large vorticities in peripheral plasmas significantly affect the intrinsic properties of the QGP, particularly its viscosity?

The results on the centrality dependence of jet quenching reported by the STAR collaboration probe, in fact, precisely the domain in the quark matter phase diagram where large vorticities have been observed in peripheral plasmas in the course of a Beam Energy Scan \cite{kn:tlusty}: that is, the domain of relatively low temperatures and relatively large values of the baryonic chemical potential. (The impact energies discussed in \cite{kn:beamenergyjet} are $\sqrt{s_{NN}} = 7.7, 11.5, 14.5, 19.6, 27, 39,$ and $62.4$ GeV.) Thus if indeed vorticity affects jet quenching, we can expect this to appear in near-future scans of the strong coupling regime\footnote{The detection of jet quenching in peripheral collisions is an extremely complicated matter, involving numerous competing effects, some of which deplete, but others enhance, the numbers of high-$p_T$ hadrons in the final state; furthermore, there are questions regarding biases associated with event selection and collision geometry.  See \cite{kn:beamenergyjet,kn:alicemod} for clear discussions of these complexities. For these reasons, it would be premature to attempt to find signs of our predictions regarding viscosity in the current, rather preliminary, observational data.}.

This is however a difficult regime from a theoretical point of view; it is one in which the methods of the gauge-gravity or \emph{holographic} \cite{kn:nat,kn:veron,kn:wolf} duality are often suggestive.

Using such methods, we have put forward \cite{kn:93,kn:95} a gauge-gravity account of QGP vorticity. In this work we propose to use this model to study the effects of vorticity on the QGP viscosity. In doing so, one must be aware of the intrinsic limitations of the gauge-gravity approach (the extent of the similarity between the boundary field theory and actual QCD depends on the circumstances and is always open to doubt) as well as the fact that many of the relevant observational parameters are not well constrained and in any case evolve rapidly with time. Nevertheless the effect we find is so striking that it merits attention: we find that our holographic model predicts that large vorticities, on the order of those actually observed by the STAR collaboration, strongly \emph{reduce} $\eta$, the shear viscosity of the QGP\footnote{One can argue \cite{kn:90} that the proper measure of ``viscosity'' is a dimensionless quantity, the \emph{Reynolds number} $Re$ (which is small for very viscous fluids): it takes into account the size, density, and characteristic velocity of the system, as well as $\eta$; $Re$ gives, for example, a way of assessing whether turbulence is likely to develop. (See \cite{kn:KelvinHelm} for a study of the Reynolds number of the QGP.) For our purposes, we can regard the size, density, and characteristic velocity as being approximately fixed, so that a downward variation in $\eta$ can be interpreted directly as an upward variation of $Re$.}, by a factor of order 4. Within the gauge-gravity approach, this means that ``peripheral'' plasmas are less viscous than their ``central'' counterparts, by a substantial margin: \emph{the two forms of the QGP are not similar} in this sense. As explained earlier, this could have consequences for many important phenomena associated with the QGP, including the centrality dependence of jet quenching.

Our model's most striking feature is a \emph{holographic bound on QGP vorticity} \cite{kn:93,kn:95}, a bound which is strongly supported by the fact that it predicts that vorticity should be \emph{smaller} (at given centrality), and consequently less easily detected, at high rather than low impact energies: a seemingly counterintuitive prediction which has nevertheless been strikingly confirmed by the observations \cite{kn:STARcoll,kn:STARcoll2,kn:STARcoll3,kn:STARcoll4}. It is also remarkable that the bound is not only \emph{respected} for a wide range of impact energies ($\sqrt{s_{\m{NN}}} = 11.5 - 200$ GeV), it is actually close to being \emph{saturated} in those cases. In this work, we find that the extent of the suppression of viscosity by vorticity (that is, the ratio of the viscosities in the peripheral and central cases) is essentially independent of the impact energy in this range. It turns out in fact that the suppression depends, in the range where vorticity has been observed, not so much on the vorticity itself ---$\,$ which varies strongly with impact energy ---$\,$ but rather on \emph{how close the vorticity of a given plasma is to saturating the bound}.

We remarked earlier that a gauge-gravity approach is appropriate here because we are dealing with extremely strongly coupled plasmas. One sign of that is that these plasmas have a large baryonic chemical potential, $\mu_B$. In a gauge-gravity approach, variations of $\mu_B$ can mimic the effects of vorticity, in a sense we will explain. Since the relevant plasmas exhibit both large vorticities and large values of the baryonic chemical potential, it is clearly necessary here to clarify the extent of this effect: so we begin with that.

\addtocounter{section}{1}
\section* {\large{\textsf{2. Holography of the Baryonic Chemical Potential}}}
In the gauge-gravity approach, the QGP is represented by certain thermal field theory in four dimensions, dual to a ``hot'' black hole in a five-dimensional asymptotically AdS bulk spacetime. For simplicity, in this work we consider only Einstein gravity in the bulk. This means \cite{kn:nat} that, at very strong coupling, the well-known ratio $\eta/s$ \cite{kn:KSS,kn:SS}, where $s$ is the entropy density of the boundary theory, is fixed. This allows us to work with $s$ instead of the viscosity itself\footnote{If one considers the (very) much more complicated higher-derivative gravity theories in the bulk, then it becomes possible to allow $\eta/s$ to vary, though only perturbatively: see \cite{kn:myers}. We will return to this issue in the Conclusion.}. That is, we assume that if $s$ is (say) reduced by a certain factor, this means that $\eta$ is likewise reduced by approximately that factor. In fact, from a holographic point of view, it is more natural to consider the \emph{specific} entropy density $s/\varepsilon$, that is, the entropy density divided by the energy density. Again, if we fix $\varepsilon$, then variations of the specific entropy density reflect similar variations of $\eta$ in the holographic picture.

Our objective in this section, then, is to understand the consequences of varying the baryonic chemical potential for $s/\varepsilon$.

In the presence of a non-trivial baryonic chemical potential, our black hole must have an electric charge\footnote{The reader can take ``non-trivial'' to mean that $\mu_B$ is of the same order as the temperature. For example, collisions of gold nuclei at $\sqrt{s_{\m{NN}}} = 27$ GeV give rise to a plasma with a temperature of around 172 MeV, and with $\mu_B \approx 150$ MeV. By contrast, the plasmas produced in collisions at 200 GeV and higher have baryonic chemical potentials which are negligible relative to the temperature.}. Leaving the effects of vorticity to one side for the present, we therefore need a five-dimensional Reissner-Nordstr\"om-AdS$_5$ black hole with a spherical event horizon. (Other topologies are possible, but ultimately we want to consider particles rotating in an equatorial plane around a rotating black hole, so we need a spherical topology here.) The metric is therefore
\begin{flalign}\label{ALEPH}
g(\m{RNAdS}_5)\;=\;&-\,\left({r^2\over L^2}\,+\,1\,-\,{2M\over r^2}\,+\,{Q^2\over 4\pi r^4}\right)\m{d}t^2\,+{\m{d}r^2\over {r^2\over L^2}\,+\,1\,-\,{2M\over r^2}\,+\,{Q^2\over 4\pi r^4}}\\ \notag \,\,\,\,&\,+\,r^2\left(\m{d}\theta^2 \,+\, \sin^2\theta\,\m{d}\phi^2\,+\,\cos^2\theta\,\m{d}\psi^2\right).
\end{flalign}
Here $t$ and $r$ are as usual, and ($\theta,\,\phi,\,\psi$) are (Hopf) coordinates on a three-dimensional sphere. We see that the spacetime is characterized by three parameters describing the geometry: $L$, $M$, and $Q$. These three parameters have units (in natural, not Planck, units) of length, squared length, and squared length respectively.

All three of these geometric parameters have physical interpretations in the bulk. Of course, $L$ describes underlying curvature of the spacetime, the curvature it would have in the absence of the black hole: that is, $L$ governs the behaviour of gravitation under those circumstances. On the other hand, $M$ and $Q$ are related to the physical mass and charge of the black hole, $m$ and $q$, by
\begin{equation}\label{BET}
m\;=\;{3\pi M\over 4\ell^3_{\mathcal{B}}},\;\;\;\;\;q\;=\;{\sqrt{3}\pi Q\over 2\ell^{3/2}_{\mathcal{B}}},
\end{equation}
where $\ell_{\mathcal{B}}$ is the gravitational length scale in the bulk. (Here the notation is chosen to be consistent with that of \cite{kn:gibperry}, since we use that reference for the five-dimensional AdS-Kerr geometry in the following Section of this work.) Since $M$ and $Q$ have units of squared length, it follows that $m$ and $q$ have the correct physical units, that is, inverse length for $m$, and (length)$^{1/2}$ for $q$. (Recall that electric charge is (in natural units) dimensionless only in a four-dimensional spacetime. In $n$ spacetime dimensions it has the units of (length)$^{(n-4)/2}$ in natural units.)

The Hawking temperature of this black hole is given by the equation
\begin{equation}\label{GIMEL}
4\pi T\;=\;{8M\over r_H^3}\,-\,{2\over r_H}\,-\,{3Q^2\over 2\pi r_H^5},
\end{equation}
where $r_H$ is the location of the outer event horizon, expressed in terms of $L$, $M$, and $Q$ by taking the larger root of
\begin{equation}\label{DALET}
{r_H^2\over L^2}\,+\,1\,-\,{2M\over r_H^2}\,+\,{Q^2\over 4\pi r_H^4}\;=\;0.
\end{equation}
The black hole entropy is
\begin{equation}\label{HE}
S\;=\;{\pi^2r_H^3\over 2\ell^3_{\mathcal{B}}}.
\end{equation}

The electromagnetic potential form is
\begin{equation}\label{VAV}
\mathbb{A}\;=\;{\sqrt{3}Q\over 8\pi \ell^{3/2}_{\mathcal{B}}}\left({1\over r^2}\,-\,{1\over r^2_H}\right)\m{d}t;
\end{equation}
the constant term has been fixed, as usual, by requiring regularity at the event horizon.

The holographic dictionary is now constructed as follows.

As is standard practice, the Hawking temperature is identified with the temperature, $T_{\infty}$, of the field theory ``at infinity'':
\begin{equation}\label{ZAYIN}
4\pi T_{\infty}\;=\;{8M\over r_H^3}\,-\,{2\over r_H}\,-\,{3Q^2\over 2\pi r_H^5},
\end{equation}
where now $r_H$ is \emph{defined} as the function of $M$, $Q$, and $L$ given by solving (\ref{DALET}).

We can identify the entropy per unit mass of the bulk black hole (the ratio of the right sides of equations (\ref{HE}) and the first entry in (\ref{BET})) with the ratio of the entropy density $s$ to the energy density $\varepsilon$ of the boundary plasma:
\begin{equation}\label{ALPHA}
{s\over \varepsilon}\;=\;{2\pi r^3_H\over 3M}.
\end{equation}
This quantity is the main focus of our attention, since it allows a holographic computation of $s$ when the energy density is known. Note that $\ell_{\mathcal{B}}$ does not appear here, just as it does not appear in the expression for the temperature; though it does appear in the expressions for the entropy and the mass of the black hole separately.

Next, the baryonic chemical potential of the plasma is to be identified \cite{kn:clifford} with ($- 3$ times) a certain multiple of the limiting value of the coefficient of $\m{d}t$ in the electromagnetic potential vector. That one needs to take a \emph{multiple} is clear from the fact that the coefficient itself has the wrong units: it has the units of an electric potential in a five-dimensional spacetime, that is, (length)$^{-3/2}$, whereas $\mu_B$ has units of (length)$^{-1}$. In other words, as was pointed out (in a different formulation) in \cite{kn:myers}, there is implicitly an additional length scale, denoted $L_*$ in \cite{kn:myers}, in this system. The baryonic chemical potential is obtained by multiplying the limiting value of the coefficient by $- 3 L_*^{1/2}$: we have
\begin{equation}\label{BETA}
\mu_B\;=\;{3 \sqrt{3} Q L_*^{1/2}\over 8 \pi r_H^2 \ell_{\mathcal{B}}^{3/2}}.
\end{equation}

In \cite{kn:myers}, $L_*$ is expressed as a multiple of $L$, and we do likewise by setting $L_* = \varpi_* L$, where $\varpi_*$ is dimensionless. (In \cite{kn:myers}, $\varpi_*$ is fixed for definiteness at $\varpi_* = \pi$, but it is hard to justify a specific choice on theoretical grounds: we regard $\varpi_*$ as a new parameter, to be fixed by data.)

We can eliminate $\ell_{\mathcal{B}}$ in (\ref{BETA}) by means of the fundamental holographic relation $\ell_{\mathcal{B}}^3/L^3 = \pi/(2N_c^2)$ (see \cite{kn:nat}), where $N_c$ is the number of colours describing the boundary field theory. It is convenient to combine all of the various dimensionless numerical parameters into one: we set $\varpi = 3\sqrt{3 \varpi_*} N_c/4\sqrt{2}\pi^{3/2}$, and so
\begin{equation}\label{GAMMA}
\mu_B\;=\;{\varpi Q\over r_H^2 L}.
\end{equation}
Here $\varpi$ is a dimensionless quantity, characteristic of the boundary field theory, in principle (but not in practice) calculable from that theory. Again, in the application to the actual QGP, it is a parameter to be fixed by data. If we fix the parameters of the field theory, then $\varpi$ must have a fixed value; concretely, we will take this to mean that $\varpi$ is fixed by the impact energy (when $\mu_B$ is not negligible).

This discussion supplies the dual interpretations of $M$ and $Q$ in terms of boundary physics. All that remains is to establish the dual interpretation of the third bulk parameter, $L$. (As above, we regard $\ell_{\mathcal{B}}$ as a ``known'' multiple of $L$.)

As the bulk and boundary physics are entirely equivalent, such an interpretation must exist. This might well vary from one application to another, but here we are ultimately interested in the QGP with a non-zero vorticity; and in that context, it was shown in \cite{kn:93} that $L$ does indeed have a clear interpretation on the boundary: it is a parameter that determines the \emph{radius of gyration} of the plasma sample. That is, it is a measure of the manner in which the plasma is distributed around the axis of rotation\footnote{Of course, the plasma sample is far from being a rigid body, so this interpretation is not to be taken literally. One thinks of it as a guide as to which parameters of the QGP can be expected to influence the value of $L$.}. In the holographic model, this radius actually depends on the angular momentum in the peripheral case (and this is how the model explains the inverse relation between angular momentum and the observed vorticities), but otherwise (and specifically for central collisions) it is independent of the collision parameters. It can be computed in the following manner.

In the context of a specific phenomenological model (for example, the colour string percolation model discussed in \cite{kn:sahoo}) one is able to compute representative values for $T_{\infty}$, $s/\varepsilon$, and $\mu_B$ for central collisions. Let us focus for the moment on central collisions at $\sqrt{s_{NN}} = 200$ GeV. As mentioned earlier, this impact energy is so high that $\mu_B$ is negligible for our purposes, so in view of equation (\ref{GAMMA}) we can set $Q = 0$. Using (from \cite{kn:sahoo}) $T_{\infty} \approx 190$ MeV (which can readily be expressed in terms of inverse femtometres, 1 fm$^{-1} \approx 197.327$ MeV), and $s/\varepsilon \approx 1.2835$ fm, we have now three equations, (\ref{DALET}),(\ref{ZAYIN}),(\ref{ALPHA}), for three variables: $M$, $r_H$, and $L$. (We will denote this last quantity by $L_0$ in this context, to indicate that this value only applies to central collisions.) We find numerically that
$L_0 \approx 0.9528$ fm.

We can now turn to the cases where $\mu_B$ is not negligible. The procedure is much as before, but now we can insert values into the left side of equation (\ref{GAMMA}), and now (with $L = L_0 \approx 0.9528$ fm) solve it, together with (\ref{DALET}),(\ref{ZAYIN}),(\ref{ALPHA}), for the four unknowns $M$, $Q$, $r_H$, and $\varpi$. All of the parameters are now fixed, and the holographic parameter correspondence is complete.

For example, consider central collisions with impact energy 27 GeV. Here the baryonic chemical potential is not negligible: from \cite{kn:sahoo} we have in this case $\mu_B \approx 150$ MeV (which agrees with the observed value given in \cite{kn:STARchem}). With (again from \cite{kn:sahoo}) $s/\varepsilon \approx 1.3347$ fm and $T_{\infty} \approx 172$ MeV, one finds, solving the four equations numerically, that $\varpi \approx 0.3821$ in this case.

With the parameters fixed, we can use these equations to determine the effect of varying one parameter on the others. In particular, we can answer the following question: how does varying $\mu_B$ (with all other parameters fixed\footnote{This variation is a purely theoretical exercise, designed to allow us to understand how one parameter depends on another. We are \emph{not} suggesting that it is experimentally possible to vary $\mu_B$, while keeping all of the other parameters fixed.}) affect the main object of our interest, namely $s/\varepsilon$?

We find that, if $\mu_B$ is decreased from its actual value, then $s/\varepsilon$ \emph{increases}; but the rate of increase is quite small. In the table, we report the values of $s/\varepsilon$ at $100 \%$, $75 \%$, $50 \%$, $25 \%$, and finally $0 \%$ of the actual value of $\mu_B$ for collisions at the three impact energies which are our main focus of attention (because they are the collisions in which QGP vorticity is most unambiguously observed).

\begin{center}
\begin{tabular}{|c|c|c|c|c|c|}
  \hline
$\sqrt{s_{\m{NN}}}$ (GeV) & $s/\varepsilon(\mu_B)$   & $s/\varepsilon(0.75\,\mu_B)$ & $s/\varepsilon(0.5\,\mu_B)$ & $s/\varepsilon(0.25\,\mu_B)$ & $s/\varepsilon(\mu_B = 0)$\\
\hline
$19.6$ &  1.3354  & 1.3717 & 1.3994 & 1.4168 & 1.4227\\
$27$  &  1.3347 & 1.3680 & 1.3934 & 1.4092 & 1.4146 \\
$39$  &  1.3202  & 1.3439  & 1.3616 & 1.3725 & 1.3762\\
\hline
\end{tabular}
\end{center}

We see that, in each case, $s/\varepsilon$ increases as $\mu_B$ is reduced; but that, even in the most extreme case where $\mu_B = 0$, the increase is only by a small factor, of around $4 \% - 6.5 \%$.

To put it another way: the effect of the baryonic chemical potential is to reduce $s/\varepsilon$, by a small amount. As we will see, vorticity has a similar but far larger effect: larger, in fact, by about two orders of magnitude. The difference is so great that, when studying the effect of vorticity on $s/\varepsilon$, we are entitled to ignore the effect of the chemical potential on that quantity.

To put it yet another way, and as a final summary of this discussion: \emph{if we neglect the effect of the baryonic chemical potential on $s/\varepsilon$, then we will over-estimate the viscosity of the plasma, by a small amount} (less than $10 \%$).

Henceforth, then, we will consider only electrically neutral five-dimensional asymptotically AdS black holes\footnote{The reader may question the need for this approximation, good though it may be: why not consider both effects simultaneously, using a five-dimensional version of the AdS-Kerr-Newman metric? The answer is that this metric \emph{has yet to be discovered}. Various approximate metrics are known when the charge or the angular momentum are small, and also in dimensions higher than five; none of these is very useful here. See \cite{kn:emparan} for a clear recent discussion of this. In any case, separate analyses are necessary in order to determine which of the two effects is dominant.}. Let us now consider those.

\addtocounter{section}{1}
\section* {\large{\textsf{3. The Holographic Dictionary in the Presence of Vorticity}}}
The bulk geometry we need is that of a five-dimensional asymptotically AdS$_5$ rotating black hole with a topologically spherical event horizon (since only these can \emph{rotate}); that is, an AdS$_5$-Kerr black hole. In the gauge-gravity duality, the physics of this spacetime is dual to that of an $\mathcal{N} = 4$ super-Yang-Mills theory defined on the four-dimensional conformal boundary, corresponding ultimately to a QGP with a non-zero angular momentum density.

We remind the reader that our immediate technical objective is to understand the possible effects of realistic values of the vorticity on the quantity $s/\varepsilon$, the ratio of the entropy and energy densities. This quantity is of interest in itself, but our primary concern is with the fact that its variations reflect, in our holographic model, variations of the shear viscosity $\eta$.

The five-dimensional AdS-Kerr spacetimes were introduced in \cite{kn:hawk}. Such black holes can rotate around two axes simultaneously, so in general one has two rotation parameters, $(a,b)$; but here we only need one axis, so we set $b = 0$. The AdS$_5$-Kerr metric in this special case takes the form
\begin{flalign}\label{A}
g(\m{AdSK}_5^{(a,0)}) = &- {\Delta_r \over \rho^2}\Bigg[\,\m{d}t \; - \; {a \over \Xi}\m{sin}^2\theta \,\m{d}\phi\Bigg]^2\;+\;{\rho^2 \over \Delta_r}\m{d}r^2\;+\;{\rho^2 \over \Delta_{\theta}}\m{d}\theta^2 \\ \notag \,\,\,\,&+\;{\m{sin}^2\theta \,\Delta_{\theta} \over \rho^2}\Bigg[a\,\m{d}t \; - \;{r^2\,+\,a^2 \over \Xi}\,\m{d}\phi\Bigg]^2 \;+\;r^2\cos^2\theta \,\m{d}\psi^2 ,
\end{flalign}
where
\begin{eqnarray}\label{B}
\rho^2& = & r^2\;+\;a^2\m{cos}^2\theta, \nonumber\\
\Delta_r & = & (r^2+a^2)\left(1 + {r^2\over L^2}\right) - 2M,\nonumber\\
\Delta_{\theta}& = & 1 - {a^2\over L^2} \, \m{cos}^2\theta, \nonumber\\
\Xi & = & 1 - {a^2\over L^2}.
\end{eqnarray}
Here ($\theta,\,\phi,\,\psi$) are Hopf coordinates on a deformed three-dimensional sphere, $L$ is the asymptotic AdS curvature length scale, $M$ and $a$ are (positive) parameters, with units (as above, in natural units) of squared length and length respectively, describing the geometry of the spacetime; they are related respectively to the black physical mass $m$ and its specific angular momentum (angular momentum per unit physical mass) $\mathcal{A}$ by \cite{kn:gibperry}
\begin{equation}\label{C}
m\;=\;{\pi M \left(2 + \Xi\right)\over 4\,\ell_{\mathcal{B}}^3\,\Xi^2},
\end{equation}
where, as before, $\ell_{\mathcal{B}}$ is the gravitational length scale in the bulk, and
\begin{equation}\label{CC}
\mathcal{A}\;=\;{2 a \over 3 - \left(a^2/L^2\right)}.
\end{equation}

The Hawking temperature is given \cite{kn:gibperry} in this case by
\begin{equation}\label{CCC}
T\;=\;{r_H\left(1 + {r_H^2\over L^2}\right)\over 2\pi \left(r_H^2 + a^2\right)} + {r_H\over 2\pi L^2},
\end{equation}
where $r_H$ denotes the horizon radius, which can be regarded as a function of $M$ and $a$ through its definition as the largest root of $\Delta_r$, that is, through
\begin{equation}\label{CROC}
(r_H^2+a^2)\left(1 + {r_H^2\over L^2}\right) - 2M \;=\;0.
\end{equation}
The entropy is
\begin{equation}\label{CCCC}
S\;=\;{\pi^2\left(r_H^2 + a^2\right)r_H\over 2\ell_{\mathcal{B}}^3\Xi}.
\end{equation}
Note that, for this bulk geometry to be well-defined, we must have \cite{kn:93}
\begin{equation}\label{D}
\mathcal{A}\;<\;a \;<\; L.
\end{equation}
These inequalities are important, for they mean that, since (as we shall shortly see) $\mathcal{A}$ increases with impact energy and has no upper bound, \emph{we cannot expect $L$ to be the same in all cases}: we have to choose $L$ separately for each impact energy (and we need to find a way of determining its value in each case).

The ``holographic dictionary'' in this case is as follows.

As always, the Hawking temperature given by (\ref{CCC}) will be identified with the temperature of the plasma-like matter in the boundary theory, $T_{\infty}$:
\begin{equation}\label{E}
T_{\infty}\;=\;{r_H\left(1 + {r_H^2\over L^2}\right)\over 2\pi \left(r_H^2 + a^2\right)} + {r_H\over 2\pi L^2}.
\end{equation}
As before, we identify the entropy per unit mass of the bulk black hole (the ratio of the right sides of equations (\ref{CCCC}) and (\ref{C})) with the ratio of the entropy density to the energy density of the boundary plasma:
\begin{equation}\label{F}
{s\over \varepsilon}\;=\;{2\pi r_H\left(r_H^2+a^2\right)\left(1 - {a^2\over L^2}\right)\, \over 3M}.
\end{equation}

The parameter $\mathcal{A}$ is interpreted on the boundary as the ratio of the angular momentum density of the plasma, $\alpha$, to its energy density\footnote{In natural units, $\alpha$ and $\varepsilon$ have dimensions respectively of fm$^{- 3}$ and fm$^{- 4}$, so $\mathcal{A}$ has units of fm.}. These densities can be estimated (using for example \cite{kn:jiang} and \cite{kn:sahoo}) and so, for a given collision of heavy ions, we can assign a definite numerical value to $\mathcal{A}$. We can then use equation (\ref{CC}) to fix the bulk geometric parameter $a$ (if we know $L$).

Once again, the final bulk parameter is $L$, and we need a dual interpretation for it that will allow us to compute it, given data in the boundary theory. As mentioned above, the inequalities (\ref{D}) strongly suggest that $L$ varies as some increasing function of $\mathcal{A}$. In the context of the holographic vorticity bound discussed in \cite{kn:93,kn:95}, an explicit relation of this sort does in fact exist; very briefly, it is as follows.

In \cite{kn:93}, we considered a plasma on the boundary with a specified ratio $\mathcal{A}$ of angular momentum to energy densities, and constructed a dual model of this system by considering matter in the bulk, also with an angular momentum to energy ratio equal to $\mathcal{A}$. Most of this matter is assumed to collapse to form a rotating black hole, but we assume also that part of it consists of particles rotating (in an equatorial orbit) with an angular velocity $\omega$: this is the dual of the QGP vorticity.

It is a lengthy but essentially elementary exercise to compute the relation between $\omega$ and $\mathcal{A}$ in the bulk system: it is
\begin{equation}\label{G}
\omega\;=\;{\mathcal{A}\over L^2}\,\sqrt{{\Xi\over 1\;+\;{\mathcal{A}^2\over L^2}\,\Xi}},
\end{equation}
where $\Xi$ is the function of $a$ and $L$ (and therefore of $\mathcal{A}$ and $L$) which is defined by the final equation in (\ref{B}), above. Thus, in the dual interpretation, the vorticity is expressed in terms of the angular momentum and energy densities and $L$. For collisions at fixed impact energy and centrality, the first two are fixed, so the vorticity is a function of $L$: this is the basis of the claim, mentioned in the previous section, that $L$ has a dual interpretation as an analogue of the radius of gyration of a rigid body (since the radius of gyration mediates between angular velocity and angular momentum).

In a collision with given impact energy and centrality, $\mathcal{A} = \alpha/\varepsilon$ is fixed, and $\omega$ becomes a function of $L$ only, given in (\ref{G}). This function is \emph{bounded above}: in other words, no matter what value $L$ takes, $\omega$ can never exceed a certain value. This is the vorticity bound: it takes the very simple form
\begin{equation}\label{H}
 \omega\;\leq\;\varkappa\;{\varepsilon\over \alpha}\,,
\end{equation}
where $\varkappa$ is a universal dimensionless constant, $\varkappa\;\approx \; 0.2782$.

One finds that, except at very low impact energies\footnote{The exceptional case is the lowest impact energy discussed in \cite{kn:STARcoll}, 7.7 GeV. However, the reported data in that case are rather hard to interpret: the difference between the average polarizations of $\Lambda$ and $\overline{\Lambda}$ hyperons is very large. It is possible that there are other effects at low impact energies which complicate the interpretation of polarization in terms of vorticity. See in this connection the recent interesting observations in \cite{kn:csernkap}.}, the observed \cite{kn:STARcoll,kn:STARcoll2} vorticities do in fact respect this bound. This is the case for nearly all of the impact energies considered in \cite{kn:STARcoll,kn:STARcoll2}, namely $\sqrt{s_{\m{NN}}} = 11.5,\, 14.5,\, 19.6,\, 27,\, 39,\, 62.4,\, 200$ GeV (at around $20 \%$ centrality).

In fact, the data do not merely respect the bound: to a good approximation, they actually \emph{saturate} it: the observed vorticities are as large as the bound permits. In other words, the value of $L$ must be such that, for given $\mathcal{A}$, $\omega$, regarded as a function of $L$, is equal to its maximum value, $\omega_{\m{max}}$. That value of $L$, denoted by $L_{\omega_{\m{max}}}$, is given (see \cite{kn:93} for the details) by
\begin{equation}\label{I}
L_{\omega_{\m{max}}}\;=\; \sigma \,\mathcal{A},
\end{equation}
where $\sigma$ is a dimensionless universal constant (determined by $\varkappa$), $\sigma \;\approx \; 1.2499$.

In this way, we can compute $L$ for given collision data in those cases in which the bound is attained\footnote{Which of course excludes near-central collisions; $L$ has to be fixed in a different way, explained in the preceding Section, in that case.}, and the holographic correspondence for the QGP with vorticity is now complete.

We are now in a position to investigate the effect of vorticity on $s/\varepsilon$, just as we investigated the effect of the baryonic chemical potential in the preceding Section.

\addtocounter{section}{1}
\section* {\large{\textsf{4. The Effect of Vorticity on the Specific Entropy Density}}}
Before we begin to study the details of the QGP produced in peripheral collisions, we should also remark on another phenomenon associated with these particular collisions: the magnetic field.

There is no doubt that the magnetic field is enormous at the very \emph{beginning} of the lifetime of a ``peripheral'' plasma \cite{kn:denghuang}. The question is whether, as a naive estimate based on the rapid departure of the spectator nucleons would suggest, this magnetic field attenuates very rapidly, and so might not have much effect over the lifetime of the plasma. This is a complicated matter which has been the subject of much discussion.

The most recent observational evidence related to this question arises in connection with the \emph{same} observations \cite{kn:STARcoll} that uncovered the existence of vorticity in the QGP: M\"{u}ller and Sch\"{a}fer observe \cite{kn:muller} that, while the \emph{sum} of the global polarization percentages of $\Lambda$ and $\overline{\Lambda}$ hyperons is sensitive to the vorticity \cite{kn:hyper}, their \emph{difference} puts a bound on the magnetic field late in the lifetime of the plasma. This bound proves to be very small (except possibly at very low impact energies, as mentioned earlier), suggesting clearly that the magnetic field does attenuate rapidly. We will therefore take it that, like the mathematically closely allied baryonic chemical potential (both arise holographically from the electromagnetic potential vector in the bulk), the magnetic field does have an effect on the viscosity, but that it is negligible relative to the effect of vorticity.

We now proceed to use the holographic model to compute the effect of vorticity on $s/\varepsilon$ for collisions at $20 \%$ centrality and impact energies $\sqrt{s_{\m{NN}}} = 11.5,\, 14.5,\, 19.6,\, 27,\, 39,\, 62.4,$ and 200 GeV. The procedure is straightforward: we use \cite{kn:sahoo} and \cite{kn:jiang} to obtain estimates for the energy densities $\varepsilon$ and the angular momentum densities $\alpha$ (see \cite{kn:93} for the details as to how this is done). Then we compute $\mathcal{A} = \alpha/\varepsilon$ and thus, using equation (\ref{I}), we find $L$ when the vorticity is maximal, $\omega = \omega_{\m{max}}$. Equation (\ref{CC}) then allows us to compute $a$, and (again using temperatures from \cite{kn:sahoo}) we then have two equations, (\ref{CROC}) and (\ref{E}), for the remaining two unknowns, $M$ and $r_H$. Finally equation (\ref{F}) gives us $s/\varepsilon$. For reasons to be explained, we also wish to repeat this calculation for the (not realistic) situation in which the vorticity bound is not saturated: to do this, we consider $\omega = 0.5 \times \omega_{\m{max}}$ (by lowering $\mathcal{A}/L$ in equation (\ref{G}) (that is, by increasing $L$), for each value of $\mathcal{A}$, until $\omega$ decreases by half). One then proceeds as before.

The table shows, for each impact energy, the value (from \cite{kn:sahoo}) of $s/\varepsilon$ for central collisions (centrality $\mathcal{C}$ = vorticity $\omega$ = 0), the value of $\mathcal{A}$ computed (see \cite{kn:93}) for collisions at that impact energy with centrality $\mathcal{C} = 20 \%$, accompanied by the value of $s/\varepsilon$ at that centrality, computed in the manner just described. We measure the effect of vorticity on viscosity by means of the ratio $V(\omega)$, defined by
\begin{equation}\label{J}
V(\omega) \equiv \frac{\displaystyle \frac{s}{\varepsilon}(\omega)}{\displaystyle \frac{s}{\varepsilon}(\omega = 0)},
\end{equation}
given in the table for $\omega = \omega_{\m{max}}$ and $\omega = 0.5 \times \omega_{\m{max}}$. The units for $\sqrt{s_{\m{NN}}}$ are, as indicated, GeV; $V(\omega)$ is of course dimensionless; for all other entries in the table, the units are femtometres.
\begin{center}
\begin{tabular}{|c|c|c|c|c|c|}
  \hline
$\sqrt{s_{\m{NN}}}$(GeV)&$\displaystyle \frac{s}{\varepsilon}(\mathcal{C}=0\%)$&$\mathcal{A}(\mathcal{C}=20 \%)$&$\displaystyle \frac{s}{\varepsilon}(\omega = \omega_{\m{max}} )$&$V(\omega_{\m{max}})$&$V(0.5 \times \omega_{\m{max}})$ \\
\hline
$11.5$ & 1.3440  &10.980 &0.3438 & 0.2558  & 0.8086\\
$14.5$ &  1.3409  &12.053 &0.3360 & 0.2506 &0.7921\\
$19.6$ &  1.3354  &13.250 &0.3305  &0.2475 &0.7821\\
$27$  &  1.3347 & 17.368 &0.3282   &0.2459 &0.7772\\
$39$  &  1.3202  & 20.386 &0.3177   &0.2406 &0.7605\\
$62.4$  &  1.3167  &31.146 &0.3156   &0.2397  &0.7575\\
$200$  &  1.2835  &71.848 &0.2971    &0.2315 &0.7316\\
\hline
\end{tabular}
\end{center}

There are two main points to be made here. First, of course, we are struck by the size of the effect: realistic (observed) values of the QGP vorticity suffice to reduce $s/\varepsilon$ by a large factor, of order 4. Furthermore, this is an underestimate: had we included the effect of the baryonic chemical potentials of these plasmas (other than those produced at the highest impact energies, which have the smallest values of $\mu_B$), the factor would be, as we know from the preceding Section, (a little) larger. Within the holographic model, this means that \emph{the shear viscosities of ``peripheral'' plasmas are very substantially smaller than those of their ``central'' counterparts}.

The second observation is that the extent of this reduction, by a factor of about 4, is \emph{almost entirely independent} of the impact energy; there is a decrease in $V(\omega_{\m{max}})$ with increasing impact energy, but the rate of decrease is essentially negligible. It follows that the extent of the effect does not depend significantly on the magnitude of the vorticity itself: in passing from $\sqrt{s_{\m{NN}}} = 11.5$ to $\sqrt{s_{\m{NN}}} = 200$ GeV, the vorticity decreases (using the vorticity bound (\ref{H}) and the data on $\mathcal{A}$ in the table) by a factor of about 6.5, yet $V(\omega_{\m{max}})$ decreases only from 0.2558 to 0.2315. One can also show that $V(\omega_{\m{max}})$ depends only weakly on the centrality, in the range where data have been taken hitherto; there is essentially no difference between $\mathcal{C} = 20 \%$ and $\mathcal{C} = 40 \%$. In fact, the only variable on which $V(\omega)$ does depend significantly is the extent to which the vorticity bound is saturated: we see from the table that artificially reducing $\omega$ to half its observed value would greatly reduce the effect of vorticity on viscosity (though not to the extent that the effect is negligible).

This second observation may have the following interesting consequence. Evidence of vorticity has been sought, but \emph{not} found \cite{kn:bed}, in heavy ion collisions at $\sqrt{s_{\m{NN}}} = 2.76$ TeV observed in the ALICE detector at the LHC. This is in accord with the vorticity bound, since $\alpha$ is of course very large at 20$\%$ centrality in those collisions. However, since the saturation of the bound is observed to improve as impact energies are increased \cite{kn:93}, it is probable that the bound is saturated at $\sqrt{s_{\m{NN}}} = 2.76$ TeV. \emph{If} it continues to be the case that the suppression of viscosity by vorticity depends primarily on the extent to which the bound is saturated, rather than on the vorticity itself, we would expect reductions of the viscosity in peripheral relative to central collisions by a substantial factor at such impact energies, even though the underlying vorticity cannot be directly observed\footnote{In these collisions, the vorticities are \emph{relatively} small, compared to those which have been deduced from the STAR observations; however, if the vorticity bound is saturated, this still means that they are on the order of  $2 \times 10^{20}\, \cdot \,$s$^{-1}$, that is, not altogether negligible.}.

In summary, vorticity tends to reduce viscosity; the extent of the reduction depends (at least in the range of impact energies such that vorticity has actually been observed) not on the actual value of the vorticity itself, but rather on how near it comes to its theoretically predicted maximum possible value as given by the vorticity bound, inequality (\ref{H}).

\addtocounter{section}{1}
\section* {\large{\textsf{5. Conclusion}}}
The recent direct observations of QGP vorticity \cite{kn:STARcoll,kn:STARcoll2} have revealed what might be regarded as a new form of the QGP: the ``vortical QGP''. The question is whether these plasmas differ substantially from their better-known counterparts generated by nearly central collisions.

We have argued here, using a holographic or gauge-gravity model, that the vortical plasmas do differ very substantially from the non-vortical QGP: in particular, they are characterised by a far smaller entropy density to energy density ratio, and therefore, within the simplest holographic models, by a far smaller (by a factor of about 4) shear viscosity.

The underlying physics of this phenomenon can only be a matter of conjecture. We may note, however, that it has been argued \cite{kn:shub} that strong magnetic fields in the QGP (at LHC rather than RHIC impact energies) constrict the relevant phase space in a specific manner, and so tend to reduce the entropy density of the plasma. It may well be that some analogous mechanism is at work here, since vorticity and magnetic fields often have similar effects\footnote{It is argued in \cite{kn:shub} that magnetic fields might also reduce $\varepsilon$ to a small extent; if the analogy between vorticity and magnetic fields continues to hold, then this effect means that the reduction of $s/\varepsilon$ implies that vorticity induces a slightly larger decrease in the entropy density $s$ than we have computed here.}.

Changes in the viscosity of the QGP should be expected to influence many of its observed properties. One example of a (potentially) observable consequence of a decrease in the shear viscosity is that it could affect the centrality dependence of jet quenching, as discussed at the beginning of this work. As mentioned earlier, there are serious difficulties in interpreting the current data in this area (see \cite{kn:beamenergyjet,kn:alicemod} and their references), but there is reason to hope that the situation will become clearer as the Beam Energy Scans progress.

The principal quantity describing jet quenching is the \emph{jet quenching parameter} $\hat{q}$, the mean squared transverse momentum acquired by a hard parton per unit distance travelled through the plasma. In holographic models \cite{kn:hong} this quantity is connected to the viscosity indirectly, through the entropy density $s\,$: one finds (see equation (7.13) of \cite{kn:hong}) that, in any conformal field theory with a gravity dual, $\hat{q}$ scales, \emph{not} with $s$, but rather with its square root. Assuming that plasmas produced at moderate centralities (such as 20$\%$, the case on which we have focused here) have energy densities which do not differ greatly from those of central plasmas, and that accounting for non-conformality does not greatly affect the ratios (see \cite{kn:hong}), this means that our model predicts that $\hat{q}$ is roughly halved in passing from central to peripheral collisions.

Of course, holographic models cannot be expected to generate precise predictions. It is best to state the case by asserting that the holographic model suggests that some significant, potentially observable, decrease in $\hat{q}$ may occur in peripheral collisions, when compared to their central counterparts.

In this work, we have followed the usual practice of using only Einstein gravity in the bulk; this means that the ratio $\eta/s$ has a fixed value. In reality, this ratio is not a constant: it is thought to vary slowly with temperature, for example \cite{kn:QGPparameters}. A variation of $\eta/s$ with temperature (or baryonic chemical potential) would not in itself affect our results here, but it suggests the following possibility: it may be that $\eta/s$ varies with vorticity. If that were the case ---$\,$ there is no reason currently, either experimental or theoretical, to indicate that this is so ---$\,$ then our results here might have to be modified. One could begin to investigate this possibility (which would in itself be of considerable interest) from a holographic point of view by following \cite{kn:myers} and considering higher-derivative gravity in the bulk. This approach is necessarily limited (by self-consistency) to perturbative variations too small to counteract the effect we have discussed here, but it might give some useful indications as to how rapidly $\eta/s$ can be expected to vary with vorticity, if at all, in the actual QGP.

\addtocounter{section}{1}
\section*{\large{\textsf{Acknowledgements}}}
The author thanks Dr Soon Wanmei for valuable discussions.

\end{document}